\documentclass[journal]{IEEEtran}

\ifCLASSINFOpdf
\else
   \usepackage[dvips]{graphicx}
\fi
\usepackage{url}

\usepackage{amssymb,amsmath,amsthm,amsfonts,amsgen,amsbsy,mathrsfs,esint}
\usepackage{hyperref, url}
\usepackage{stfloats}
\usepackage{graphicx}
\usepackage{cite}
\usepackage{bm}
\usepackage[utf8]{inputenc}
\usepackage{tikz}
\usepackage{float}
\usepackage{algorithmic,algorithm}
\usepackage{multicol,multirow,array}
\usepackage{enumerate}
\usepackage{booktabs}
\usepackage{siunitx}
\usepackage{setspace}
\usepackage{bbm}
\usepackage{color}
\usepackage{epstopdf}
\usepackage{ragged2e}

\newcommand{\mN}{\mathcal{N}}
\newcommand{\mJ}{\mathcal{J}}
\newcommand{\mT}{\mathcal{T}}

\newcommand{\bx}{\mathbf{x}}
\newcommand{\bz}{\mathbf{z}}
\newcommand{\bQ}{\mathbf{Q}}

\begin{document}

\title{Population Monte Carlo with Normalizing Flow}

\author{Soumyasundar Pal, Antonios Valkanas, and Mark Coates
\thanks{Soumyasundar Pal is with Huawei Noah's Arc Lab, Montr{\'e}al, Qu{\'e}bec, Canada.
Antonios Valkanas and Mark Coates are with the Dept. of Electrical and Computer Engineering, McGill University, Montr{\'e}al, Qu{\'e}bec, Canada. Corresponding author: Soumyasundar Pal. 
(e-mail: soumyasundarpal@gmail.com; antonios.valkanas@mail.mcgill.ca; mark.coates@mcgill.ca)\\
Code to reproduce our experiments is available at \url{https://github.com/soumyapal91/AIS_flow}}
}

\maketitle
\begin{abstract}
Adaptive importance sampling (AIS) methods provide a useful alternative to Markov Chain Monte Carlo (MCMC) algorithms for performing inference of intractable distributions. 
Population Monte Carlo (PMC) algorithms constitute a family of AIS approaches which adapt the proposal distributions iteratively to improve the approximation of the target distribution.
Recent work in this area primarily focuses on ameliorating the proposal adaptation procedure for high-dimensional applications.
However, most of the AIS algorithms use simple proposal distributions for sampling, which might be inadequate in exploring target distributions with intricate geometries.
In this work, we construct expressive proposal distributions in the AIS framework using normalizing flow, an appealing approach for modeling complex distributions.
We use an iterative parameter update rule to enhance the approximation of the target distribution.
Numerical experiments show that in high-dimensional settings, the proposed algorithm offers significantly improved performance compared to the existing techniques.
\end{abstract}

\vspace{-0.75cm}
\section{Introduction}
Monte Carlo (MC) methods~\cite{robert2004, liu2004} have numerous applications in statistical signal processing, communications, and machine learning. For example, in the Bayesian setting, the statistical inference of an unknown parameter is carried out by computing the posterior distribution conditioned on the observed data~\cite{candy2016}. However, in many practical problems of interest, the posterior distribution does not have an analytic form, and approximation via MC methods becomes necessary.

Markov Chain Monte Carlo (MCMC)~\cite{gilks1995} algorithms have been a popular approach for addressing posterior inference problems. They involve constructing a Markov process whose stationary distribution is the same as the target posterior. Several efficient MCMC algorithms~\cite{neal2011, girolami2011} have been developed which obtain state-of-the-art results in many high dimensional scenarios. Adaptive versions of MCMC algorithms~\cite{roberts2009, mahendran2012} are often used for accelerating the mixing of the Markov chain. Despite the effectiveness of the MCMC techniques, there are several disadvantages of these algorithms. In higher dimensions, they often result in very high computational costs and/or slow mixing. In this setting, an importance sampling (IS) based methodology can provide an alternative approach.

Importance sampling is a class of Monte Carlo methods, where the target distribution is approximated by the weighted samples from another proposal distribution~\cite{robert2004}. In high-dimensional applications, the design of an effective proposal distribution is crucial since a greater mismatch between the target and the proposal distributions leads to a high variance of the IS estimator. Several proposal distributions can be used for efficient exploration of state space of complicated target distributions; this approach is referred to as multiple IS (MIS)~\cite{elvira2015b}. 

Although the use of multiple proposals can be beneficial in some cases, adaptation of the proposal distributions~\cite{bugallo2017} can be more advantageous for inference of higher dimensional target distributions. In adaptive IS (AIS) schemes, iterative adaptation of the proposal distributions is performed to improve the approximation of the target distribution. Generally, an iteration of an AIS scheme is composed of three different steps~\cite{bugallo2017}: a) sampling from a set of proposals (sampling), b) computing
importance weights for the samples (weighting), and c) updating
(adapting) the parameters of the proposals for the next iteration.

There are numerous variants of the AIS algorithms, and the differences among them arise from consideration of how the three aforementioned steps are implemented. 
For example, the standard Population Monte Carlo (PMC) algorithm~\cite{cappe2004} draws only one sample from each proposal at each iteration, whereas the more recent schemes~\cite{elvira2017,elvira2019,elvira2015, laham2019} use multiple samples. 
The weighting of the samples is not uniform in the literature either.~\cite{cappe2004} uses individual proposals in the denominator of the importance weight computation step, whereas~\cite{elvira2017,elvira2019} use a deterministic mixture (DM-PMC) to compute the weights, as this leads to lower variance estimates (albeit at the cost of additional complexity).
A detailed comparison of related algorithms in terms of their similarities and differences is provided in~\cite{bugallo2017}.

From a practical point of view, the most crucial step in an AIS algorithm is proposal adaptation. Earlier works like~\cite{cappe2004} and its variants~\cite{elvira2017,elvira2017b} use resampling of the samples to adapt the location parameters of the proposal distributions, which is inefficient in higher dimensions. More effective resampling strategies are considered in~\cite{elvira2017}. Intuitively the goal of the adaptation step is to move the proposals in a way such that they are placed in the high-density regions of the target distribution. One way to achieve this is to use the gradient of the target to adapt the location parameters of the proposals~\cite{elvira2015,elvira2019,elvira2022,elvira2023}, although other strategies like minimizing KL divergence~\cite{cappe2008}, employing MCMC steps~\cite{martino2017,mousavi2021,llorente2022}, performing moment matching~\cite{cornuet2012} or using stochastic optimization~\cite{laham2019} can also be used. While these techniques offer improvement in performance, the application of AIS algorithms in high dimensions remains challenging.

A major limitation of the existing AIS algorithms is their use of simple proposal distributions. 
Most of these methods use isotropic Gaussian distributions as proposals for ease of sampling and density evaluation. Despite the presence of an effective adaptation rule, such a choice might be sub-optimal for the traversal of high-dimensional target distributions with multi-modality, complex correlation structures, and highly non-uniform scales across different dimensions. 
Some existing methods~\cite{elvira2019,elvira2022} address this issue to some extent by updating the scale parameters of the proposals using the local curvature of the target, along with the locations.
However, such strategies scale poorly to high dimensions since they require the computation of the inverse of the Hessian matrix of the logarithm of the target distribution. 

Normalizing Flows (NF)~\cite{rezende2015, dinh2017,papamakarios2017} represent a class of expressive probability distributions, which enables efficient and exact sampling as well as density evaluation. Various flow models have widespread applications in generative modelling~\cite{rezende2015,kingma2016}, density estimation~\cite{germain2015,dinh2017,papamakarios2017}, and modelling conditional distributions~\cite{trippe2018,winkler2019,lu2020}. Several approaches~\cite{midgley2023,parno2018,arbel2021,gabrie2022,grumitt2022} combine NF models with annealed importance sampling and MCMC for efficient sampling from intractable distributions.

In this paper, we present a novel PMC algorithm, which uses normalizing flow to parameterize effective proposal distributions. The incorporation of normalizing flow enhances the sampling capability of the PMC algorithm remarkably which results in impressive performance in high dimensions with favorable complexity. 

The paper is organized as follows.  Sec.~\ref{sec:ps} states the inference task that we address. Sec.~\ref{sec:background} reviews the AIS framework and normalizing flow. Sec.~\ref{sec:prop} introduces the proposed PMC algorithm using normalizing flow and Sec.~\ref{sec:experiments} presents and discusses the results of numerical simulation experiments. Concluding remarks are summarized in Sec.~\ref{sec:conclusion}. 
\vspace{-0.25cm}
\section{Problem Statement}
\label{sec:ps}
Although Monte Carlo methods can be applied for sampling from any probability distribution, in practice they are often used to target inference of posterior distributions in a Bayesian setting.
Let us consider the problem of statistical inference of an unknown parameter $\bx \in \mathbb{R}^d$ in the presence of some observed data $\mathbf{z} \in \mathbb{R}^{d_z}$. The prior distribution of $\bx$ is denoted by $p(\bx)$ and the observed data $\mathbf{z}$ is related to parameter $\bx$ through a probabilistic generative model, specified by the likelihood function $p(\mathbf{z}|\bx)$. 
The posterior distribution of $\bx$ is computed using Bayes' rule:
\begin{align}
\tilde{\pi}(\bx|\mathbf{z}) = \frac{p(\bx)p(\mathbf{z}|\bx)}{p(\mathbf{z})} \propto \pi(\bx|\mathbf{z})\,.\label{eq:bayes}
\end{align}
Here, $\pi(\bx|\bz){=}p(\bx)p(\mathbf{z}|\bx)$ is the unnormalized posterior distribution of $\bx$ and $p(\mathbf{z}){=}\int p(\bx)p(\mathbf{z}|\bx) d\bx >0$ is a non-negative function of $\mathbf{z}$, which is typically unknown.
So, in general, the posterior distribution $\tilde{\pi}(\bx|\mathbf{z})$ is known only up to a normalizing constant. 
In many problems, Bayesian decision-making involves computing the expectation of a function $f$ w.r.t. the posterior distribution, given as
\begin{align}
\mathbf{E}_{\tilde{\pi}}[f] = \int f(\bx)\tilde{\pi}(\bx|\mathbf{z}) d\bx = \frac{1}{p(\mathbf{z})}\int f(\bx)\pi(\bx|\mathbf{z}) d\bx\,.\label{eq:integral}
\end{align}
Here $f$ is an integrable function with respect to $\tilde{\pi}(\bx|\bz)$.
However, in most cases the integral in Eq.~\eqref{eq:integral} cannot be computed analytically because $\pi$ and/or $f$  are complicated and/or $p(\bz)$ is unknown.
In these situations, our goal is to approximate the posterior distribution by its samples. 
From here onward, we denote the target (posterior) distribution as $\pi(\bx)$ to simplify notation.
\vspace{-0.25cm}
\section{Background}
\label{sec:background}
\subsection{Generic AIS Algorithm}
Let $\pi(\mathbf{x})$ be the target distribution with a possibly unknown normalizing constant.
We assume that there are $N$ proposal distributions, used at each of $J$ total iterations of the algorithm.
Let $q_n(\cdot; \theta_n^j)$ be the $n$-th proposal at the $j$-th iteration, where $\theta_{n}^j$ typically consists of a location parameter $\mu_{n}^j$ and a scale parameter $\Sigma_{n}^j$.
In most AIS algorithms, Gaussian proposals are used for sampling, which implies that $\mu_{n}^j$ and $\Sigma_{n}^j$ are the mean vector and the covariance matrix of $q_n$ respectively.
In each iteration, we draw $K$ samples $\{\mathbf{x}_{n,j}^{(k)}\}_{k{=}1}^K{\sim}q_n(\cdot; \theta_n^j)$ from each of the proposals.
The generic AIS~\cite{bugallo2017} method is summarized in Algorithm~\ref{alg:ais}.
\begin{algorithm}[ht]
\caption{Generic AIS algorithm}
\label{alg:ais}
\begin{algorithmic}[1]
\STATE {\bfseries Input:}  $\pi(\mathbf{x})$, $N, J, K$, $\{\theta_n^1\}_{n=1}^N$
\STATE {\bfseries Output:}  $\{\omega_{n,j}^{(k)},\mathbf{x}_{n,j}^{(k)}\}, n{=}1:N, k{=}1:K, j{=}1:J$
\vspace{0.1cm}
\FOR{$j{=}1:J$}
\STATE Sampling: draw $K$ samples from each of $N$ proposals, $\{x_{n,j}^{(k)}\}_{k{=}1}^K {\sim} q_n(\cdot; \theta_n^j)$

\STATE Weighting: compute importance weight $\omega_{n,j}^{(k)}$ of $\bx_{n,j}^{(k)}$ 

\STATE Adaptation: update the proposal parameters: $\{\theta_n^j\}_{n{=}1}^N \rightarrow \{\theta_n^{j+1}\}_{n{=}1}^N$
\ENDFOR
\end{algorithmic}
\end{algorithm}
\vspace{-0.5cm}
\subsection{Normalizing Flow}
Normalizing Flow (NF)~\cite{rezende2015,dinh2017,papamakarios2017} approaches are capable of constructing complex distributions by transforming a simple probability density (which is easy to sample from and easy to evaluate) via a sequence of invertible and differentiable mappings.
These invertible transformations are designed in such a way that the forward mappings, their Jacobian matrices (and their determinants), and the inverse mappings can be computed efficiently.
The change of variable formula is used to evaluate the density of the transformed random variable.

Suppose $\bx'{\sim}q(\bx')$ is mapped to $\bx{=}\mT(\bx')$  via an invertible function $\mT^{\phi}: \mathbb{R}^d \rightarrow \mathbb{R}^d$.
Here, $\phi$ denotes the parameters of $\mT^{\phi}$.
Different NF models use different neural architectures to parameterize $\mT^{\phi}$.
The density of $\bx$ is computed in closed form as:
\begin{align}
q(\bx) = q(\bx') \lvert \text{det}(\mJ_{\mT^{\phi}}(\bx'))\rvert^{-1}\,,\label{eq:nf_density}
\end{align}
where $\mJ_{\mT^{\phi}}(\cdot) \in \mathbb{R}^{d \times d}$ is the Jacobian of the transformation and $\text{det}(\cdot)$ denotes the determinant of a matrix. 
We design $\mT^{\phi}(\cdot)$ using the real-NVP~\cite{dinh2017} architecture, composed of two coupling layers in an alternating pattern as follows:
\begin{align}
x_{\lceil \frac{d}{2} \rceil +1:d} &{=} x'_{\lceil \frac{d}{2} \rceil +1:d} \odot \exp\Big(s_{1}^{\phi_{1}}\big(x'_{1:\lceil \frac{d}{2} \rceil}\big)\Big) + t_{1}^{\phi_{2}}\big(x'_{1:\lceil \frac{d}{2} \rceil}\big)\,,\nonumber\\
x_{1:\lceil \frac{d}{2} \rceil} &= x'_{1:\lceil \frac{d}{2} \rceil} \odot \exp\Big(s_{2}^{\phi_{3}}\big(x_{\lceil \frac{d}{2} \rceil+1:d}\big)\Big) + t_{2}^{\phi_{4}}\big(x_{\lceil \frac{d}{2} \rceil+1:d})\,. 
\end{align}
Here, $\odot$ denotes Hadamard product, and $\big(s_{1}^{\phi_{1}}(\cdot), t_{1}^{\phi_{2}}(\cdot)\big)$ and $\big(s_{2}^{\phi_{3}}(\cdot), t_{2}^{\phi_{4}}(\cdot)\big)$ are the scale and translation functions of the first and second coupling layers respectively. Each of these functions is parameterized by a separate \emph{feed-forward network}~(FFN) and $\phi = \{\phi_{1:4}\}$ denotes the set of all weight matrices and bias vectors of these FFNs.
\vspace{-0.25cm}
\section{PMC with NF}
\label{sec:prop}
\vspace{-0.15cm}
\subsection{Proposal Step}
We design the $n$-th proposal distribution $q_n$ by constructing a normalizing flow $\mT_{n}^{\phi_{n}}(\cdot)$ with a multivariate normal base distribution with mean $\mu_{n}$ and covariance $\Sigma_{n}$. 
In other words, sampling from $q_n$ is carried out by first sampling $\bx'{\sim}\mN(\mu_{n}, \Sigma_{n})$ and then computing $\bx{=}\mT_{n}^{ \phi_{n}}(\bx')$. The proposal is evaluated as follows:
\begin{align}
q_{n}(\bx; \theta_{n}) = \mN(\bx'; \mu_{n}, \Sigma_{n})\lvert \text{det}(\mJ_{\mT_{n}^{\phi_{n}}}(\bx'))\rvert^{-1}\,,\label{eq:prop_n}
\end{align}
The parameters of the $n$-th proposal density are defined as $\theta_{n}{=}\{\mu_{n}, \Sigma_{n}, \phi_{n}\}$ and $\theta{=}\{\theta_{n}\}_{n{=}1}^N$ denotes all proposal parameters. Utilization of normalizing flow allows $q_n$ to be non-Gaussian and expressive. 
\vspace{-0.5cm}
\subsection{Weighting Step}
Let $\theta_n^j$ be the value of $\theta_n$ at the $j$-th iteration. 
We use the deterministic mixture (DM) weighting~\cite{elvira2017b} scheme, as it reduces the variance of the unnormalized AIS estimator, i.e., we compute:
\begin{align}
\omega_{n,j}^{(k)} \propto \frac{\pi(\bx_{n,j}^{(k)})}{\frac{1}{N}\displaystyle{\sum_{l=1}^N} q_{l}(\bx_{n,j}^{(k)}; \theta_l^j)}\,.
\end{align}
\vspace{-0.75cm}
\subsection{Adaptation Step}
We adapt the proposal parameters $\theta$ by performing a mini-batch gradient descent step towards minimizing the KL divergence between the mixture proposal distribution $q(\bx; \theta) = \frac{1}{N}\sum_{n{=}1}^N q_{n}(\bx; \theta_{n})$ and the target density $\pi(\bx)$ w.r.t. $\theta$. 
Since this KL divergence cannot be evaluated in closed form, at the $j$-th iteration, we compute an unbiased Monte Carlo estimate of $\ell(\theta^j) = \text{KL}\big(q(\bx; \theta^j)\lvert \rvert\pi(\bx)\big)$ as follows:
\begin{align}
\widehat{\ell(\theta^j)} 
&= \frac{-1}{NK} \sum_{n=1}^N \sum_{k=1}^K \log \omega_{n,j}^{(k)}\,.\label{eq:loss_approx}
\end{align}
The update rule for $\theta$ is:
\begin{align}
\theta^{j+1} = \theta^j - \epsilon^j \nabla_{\theta}\widehat{\ell(\theta)}\rvert_{\theta=\theta^j}\,,\label{eq:sgd} 
\end{align}
where $\epsilon^j$ denotes the learning rate at the $j$-th iteration. 
In practice, the stochastic gradient used in Eq.~\eqref{eq:sgd} is computed using automatic differentiation.

\vspace{-0.25cm}
\section{Numerical Experiments}
\label{sec:experiments}
We conduct simulation experiments for two scenarios.
The goal of the first experiment is to draw samples from a high-dimensional Gaussian mixture target distribution, whereas the second experiment considers a high-dimensional Bayesian logistic regression problem.
The proposed Normalizing Flow (NF)-PMC algorithm is compared with PMC~\cite{cappe2004}, Global Resampling (GR)-PMC~\cite{elvira2017}, Local Resampling (LR)-PMC~\cite{elvira2017}, Scaled Langevin (SL)-PMC~\cite{elvira2019}, Optimized (O)-PMC~\cite{elvira2022}, GRadient-based Adaptive Multiple Importance Sampler~\cite{elvira2023}, Hamiltonian Adaptive Importance Sampler (HAIS)~\cite{mousavi2021}, and Variational Adaptive Population Importance Sampler (VAPIS)~\cite{laham2019} algorithms.

GR-PMC and LR-PMC algorithms are two variants of the DM-PMC~\cite{elvira2017b} technique.
Langevin dynamics~\cite{roberts2002} is used for designing the adaptation rule of SL-PMC, whereas O-PMC and HAIS algorithms utilize Newton's optimization and Hamiltonian Monte Carlo~\cite{duane1987,neal2011} respectively. 
GRAMIS adapts the location parameters of the proposals using a physics-inspired repulsion term, which facilitates coordinated exploration of the target density.  
Location parameters of the deterministic mixture proposal distribution are adapted by minimizing the per-sample variance of the IS estimator of the normalizing constant via stochastic gradient variational inference~\cite{hoffman2013} for the VAPIS algorithm. 

For each algorithm, except PMC, we use $J{=}50$ iterations, in each of which $K{=}10$ samples are randomly drawn from each of the $N{=}100$ proposal distributions.
Since the PMC algorithm uses $K{=}1$ sample from each proposal at each iteration, we use $J{=}50\times10{=}500$ iterations for PMC, so that all of the algorithms generate the same number of samples in total.
For each setting, we conduct 100 Monte Carlo trials.

Deterministic mixture (DM)~\cite{elvira2017b} weights are computed for all algorithms except PMC, which uses the standard weighting.
As in the original papers, we follow the same backtracking strategy for tuning the step size for SL-PMC and O-PMC.
The burn-in period in HAIS is set to 10, and we use 50 leapfrog steps.  The RMSprop\footnote{\url{http://www.cs.toronto.edu/~tijmen/csc321/slides/lecture_slides_lec6.pdf}} optimizer with a decaying learning rate schedule is employed for VAPIS  and NF-PMC because of its faster convergence compared to the SGD method. 
The scale and translation functions in each coupling layer are parameterized by FFNs with 2 hidden layers with 8 neurons and $tanh$ activation for the implementation of NF-PMC. 
Additionally, we treat $\Sigma_{n}$ as hyperparameters and share the NF parameters $\{\phi_{n}\}_{n{=}1}^N$ across all proposals to reduce parameter complexity.
\begin{table*}[ht]
\scriptsize
\vspace{-0.25cm}
\caption{Mean and standard deviation of MSE in the mean estimation of the Gaussian mixture along with avg. runtime, the lowest MSE in each row is shown in bold.}
\vspace{-0.25cm}
\centering
\setlength{\tabcolsep}{5pt}
\begin{tabular}{|c|c|c|c|c|c|c|c|c|c|}
\hline 
\textbf{Alg.} &\textbf{PMC} &\textbf{GR-PMC} &\textbf{LR-PMC} &\textbf{SL-PMC} &\textbf{O-PMC} &\textbf{GRAMIS} &\textbf{HAIS} &\textbf{VAPIS} &\textbf{NF-PMC} \\ \hline
\multicolumn{9}{|c|}{\textbf{MSE}}                         \\ \hline \hline 
\textbf{$\sigma{=}1$} &36.24$\pm$5.21 &32.49$\pm$4.32             &38.65$\pm$4.62        &18.21$\pm$7.91               &17.08$\pm$7.75  &13.46$\pm$7.88  &20.71$\pm$2.80   &29.26$\pm$2.88       &\textbf{10.49$\pm$2.00}*                 \\ \hline
\textbf{$\sigma{=}2$} &61.65$\pm$14.17 &46.46$\pm$5.35 &39.72$\pm$6.82 &13.96$\pm$8.34 &59.31$\pm$10.57 &40.71$\pm$12.28 &22.60$\pm$3.05 &30.63$\pm$3.42 &\textbf{10.38$\pm$1.94}*                \\ \hline
\textbf{$\sigma{=}3$} &54.97$\pm$21.49 &67.45$\pm$7.52 &41.14$\pm$3.44 &28.86$\pm$8.19 &28.86$\pm$8.19 &28.10$\pm$8.20 &26.80$\pm$3.46 &34.11$\pm$3.69 &\textbf{10.89$\pm$1.86}*               \\ \hline
\multicolumn{9}{|c|}{\textbf{Average Runtime (seconds/iteration)}}  \\ \hline
&0.31 &0.59 &0.56 &1.92 &2.33 &2.37 &11.22 &0.70 &0.77                 \\ \hline
\end{tabular}
\label{tab:gmm}
\end{table*}

\begin{table*}[htbp]
\scriptsize
\vspace{-0.5cm}
\caption{Mean and standard deviation of relative MSE in estimating the weight vector $\bx$ and test log-likelihood in the Bayesian logistic regression along with avg. runtime, the lowest relative MSE and the highest test log-likelihood in each row are shown in bold.}
\vspace{-0.25cm}
\centering
\setlength{\tabcolsep}{6pt}
\begin{tabular}{|c|c|c|c|c|c|c|c|c|c|}
\hline
\textbf{Alg.} &\textbf{PMC} &\textbf{GR-PMC} &\textbf{LR-PMC} &\textbf{SL-PMC} &\textbf{O-PMC} &\textbf{GRAMIS} &\textbf{HAIS} &\textbf{VAPIS} &\textbf{NF-PMC} \\ \hline
\multicolumn{10}{|c|}{\textbf{Relative MSE}}                         \\ \hline \hline 
\textbf{$\sigma{=}1$} &0.72$\pm$0.18 &0.66$\pm$0.15 &0.95$\pm$0.12 &0.51$\pm$0.12 &0.48$\pm$0.13 &0.60$\pm$0.07 &0.49$\pm$0.20 &0.96$\pm$0.07 &\textbf{0.42$\pm$0.13}*                 \\ \hline
\textbf{$\sigma{=}2$} &1.35$\pm$0.24  &0.82$\pm$0.19 &1.35$\pm$0.19 &0.51$\pm$0.15 &0.46$\pm$0.14 &0.46$\pm$0.11 &0.49$\pm$0.20 &0.95$\pm$0.08 &\textbf{0.42$\pm$0.12}*             \\ \hline
\textbf{$\sigma{=}3$} &2.40$\pm$0.53 &1.32$\pm$0.24 &1.26$\pm$0.17 &0.59$\pm$0.17 &0.60$\pm$0.15 &0.47$\pm$0.11 &0.51$\pm$0.20 &0.99$\pm$0.08 &\textbf{0.41$\pm$0.11}*               \\ \hline \hline
\multicolumn{10}{|c|}{\textbf{Test Log-Likelihood}}                         \\ \hline \hline 
\textbf{$\sigma{=}1$} &-1.66$\pm$0.35 &-1.37$\pm$0.26 &-2.17$\pm$0.47 &-1.00$\pm$0.21 &-0.92$\pm$0.17 &-0.81$\pm$0.15 &-1.02$\pm$0.18 &-1.96$\pm$0.47 &\textbf{-0.75$\pm$0.16}*               \\ \hline
\textbf{$\sigma{=}2$} &-2.39$\pm$0.54 &-1.90$\pm$0.41 &-3.17$\pm$0.79 &-1.20$\pm$0.25 &-1.09$\pm$0.23 &-0.88$\pm$0.20 &-1.12$\pm$0.20 &-1.97$\pm$0.46 &\textbf{-0.76$\pm$0.15}*               \\ \hline
\textbf{$\sigma{=}3$} &-3.00$\pm$0.67 &-2.59$\pm$0.53 &-2.75$\pm$0.69 &-1.44$\pm$0.32 &-1.46$\pm$0.33 &-1.03$\pm$0.24 &-1.21$\pm$0.24  &-2.12$\pm$0.48 &\textbf{-0.74$\pm$0.15}*               \\ \hline
\multicolumn{9}{|c|}{\textbf{Average Runtime (seconds/iteration)}}  \\ \hline
&0.23 &0.61 &0.58 &1.12 &1.52 &1.56 &3.23 &0.69 &0.73                 \\ \hline
\end{tabular}
\label{tab:logistic}
\vspace{-0.5cm}
\end{table*}
\vspace{-0.5cm}
\subsection{Gaussian Mixture Target Distribution}
\label{subsec:gmm}
We consider the task of approximating a mixture of $P$ Gaussian densities in $\mathbb{R}^d$, given as $\pi(\bx){=}\sum_{p=1}^P \alpha_p \mN(\bx, \mathbf{m}_p, \bQ_p)$, where $\alpha_p, \mathbf{m}_p$, and $\bQ_p$ denote the proportion, mean, and covariance of the $p$-th mixture component.
We set $d{=}200$ and $P{=}5$. 
In each trial, a different target density is considered by sampling $(\alpha_1, \cdots, \alpha_p){\sim}Dir(10.\mathbf{1}_P)$, $\mathbf{m}_p \sim Unif([-10, 10]^{d})$, and setting $\bQ_p{=}\widetilde{\bQ}_p + 2\mathbb{I}_d$, where $\widetilde{\bQ}_p \sim \mathcal{IW}(\mathbb{I}_d, d)$. Here $Dir(10.\mathbf{1}_P)$, $Unif(\mathcal{S})$, and $\mathcal{IW}(\mathcal{B}, d_f)$ denote the Dirichlet distribution of order $P$ with parameters $(10,\cdots,10)$, uniform distribution over the set $\mathcal{S}$, and inverse Wishart distributions with scale matrix $\mathcal{B}$ and degree of freedom $d_f$ respectively. The goal is to estimate the mean of this multi-modal target distribution.

In each trial, the proposal means $\{\mu_{n}^1 \in \mathbb{R}^d\}_{n{=}1:N}$ are initialized by sampling from $Unif([-10, 10]^{d})$ for all algorithms except the proposed NF-PMC approach.
We conduct three different experiments by initializing  $\Sigma_{n}^1{=}\sigma^2\mathbf{I}_{d \times d}$ with $\sigma \in \{1,2,3\}$ for $1 \leqslant n \leqslant N$. We use the same $\{\mu_{n}^1, \Sigma_{n}^1\}$ for initializing the base distributions of the real-NVP and Xavier initialization~\cite{glorot2010} for coupling layer parameters. The initial learning rate is set to 0.25 and 0.005 for the VAPIS and the NF-PMC algorithms respectively. HAIS uses a fixed step size of 0.005.

Table~\ref{tab:gmm} shows the average MSE in the estimation of the mean of $\bx$.
The result from the best-performing approach in each experiment is marked with an asterisk if it is significantly better at the 0.5\% level compared to all other baselines based on the Wilcoxon signed-rank test.
We observe that PMC, GR-PMC, and LR-PMC algorithms struggle in this high-dimensional, multi-modal setting because of the inefficiency of resampling-based proposal adaptation.
Although SL-PMC, O-PMC, GRAMIS, HAIS, and VAPIS methods achieve improved performance, they are significantly outperformed by the proposed NF-PMC algorithm in all cases.
Moreover, NF-PMC exhibits the lowest variability of MSE for different choices of $\sigma$ among all algorithms.
\vspace{-0.5cm}
\subsection{Bayesian Logistic Regression}
\label{subsec:logistic}
We consider a logistic regression problem where the binary response $y_i \in \{0, 1\}$ is related to the predictor $\mathbf{z}_i \in \mathbb{R}^{d}$  via the following likelihood function:
\vspace{-0.125cm}
\begin{align}
p(y_i|\mathbf{z}_i, \bx) = q_i^{y_i}(1-q_i)^{(1-y_i)}, \text{ } q_i = (1+e^{-\mathbf{z}_i^T\bx})^{-1}\,\label{eq:logistic}.
\end{align}
We assume that the prior distribution of the weight vector $\bx \in \mathbb{R}^{d}$ is $\mathcal{N}(\mathbf{0}, \zeta^2\mathbf{I})$.
Conditioned on $N_{train}$ observations $(\mathbf{z}_i, y_i)$, our objective is to approximate the posterior distribution of $\bx$, which can be evaluated up to a constant as:
\vspace{-0.125cm}
\begin{align}
\tilde{\pi}(\bx| \{\mathbf{z}_i, y_i\}_{i=1}^{N_{train}}) \propto p(\bx) \prod_{i=1}^{N_{train}}p(y_i|\mathbf{z}_i, \bx)\,.
\end{align}
In each trial, the weight vector $\bx$ is sampled from its prior distribution.
The artificial predictors $\mathbf{z}_i{=}[z_{i,1}, z_{i,2},...z_{i,d}]^T$ are simulated as follows: $z_{i,1}{=}1$ and $z_{i,\ell}{\sim}\mathcal{N}(0, \delta^2)$. 
We generate $y_i$ according to the model in Eq.~\eqref{eq:logistic}. 
$N_{train}{=}500$ artificial data points are used for training and we construct a held-out test set using another $N_{test}{=}500$ points.
We set $d{=}200, \zeta{=}10$ and $\delta{=}0.1$. 
For all algorithms, we follow the same initialization strategy as in Section~\ref{subsec:gmm}. 
The initial learning rate is set to 0.5 and 0.05 for the VAPIS and the NF-PMC algorithms respectively. We use a fixed step size of 0.05 for HAIS.

In addition to reporting the relative MSE in estimating the weight vector $\bx$ as a performance metric, we estimate the average predictive log-likelihood for the held-out test data using the Monte Carlo approximation of the posterior distribution of $\bx$.
From Table~\ref{tab:logistic}, we observe that in each setting, the proposed PF-PMC algorithm obtains the lowest average relative MSE for the estimation of the weight vector and the highest average test log-likelihood among all competing techniques.

\vspace{-0.35cm}
\section{Conclusion}
\label{sec:conclusion}
\vspace{-0.15cm}
Instead of the common practice of drawing samples from isotropic Gaussian distributions in most AIS algorithms, we employ a proposal mechanism using normalizing flow and adapt its parameters by minimizing a discrepancy measure between the proposal and target distributions.
Experimental results demonstrate that the superior capability of the normalizing flow proposals in approximating complex, high-dimensional target distributions results in significantly improved performance. 
The computational burden of the proposed NF-PMC algorithm remains comparable to that of resampling-based PMC variants (e.g. GR-PMC and LR-PMC) for generating an equal number of samples.
In contrast, several other alternatives require considerably higher execution time and scale poorly to high dimensions.
An interesting future research direction is to explore minimization of the $\chi^2$ divergence instead of the KL divergence. In some settings, this has been shown to have superior sampling performance and has desirable theoretical properties~\cite{laham2019,midgley2023}. In our experiments, minimizing the KL divergence was faster, provided more stable convergence, and achieved better sampling performance.
Another potential avenue is an investigation of the applicability of other normalizing flow architectures in the AIS framework. 
For example, observation-driven proposals using conditional normalizing flow~\cite{trippe2018,winkler2019,lu2020} might be beneficial in Bayesian inference problems, whereas auto-regressive flow~\cite{germain2015,kingma2016,papamakarios2017} might be better suited for approximating high-dimensional target distributions with complex dependency structures.  

\bibliographystyle{IEEEbib}
\bibliography{refs}

\begin{thebibliography}{10}

\bibitem{robert2004}
C.~P. Robert and G.~Casella,
\newblock {\em Monte Carlo Statistical Methods},
\newblock Springer, 2004.

\bibitem{liu2004}
J.~S. Liu,
\newblock {\em Monte Carlo Strategies in Scientific Computing},
\newblock Springer, 2004.

\bibitem{candy2016}
J.~V. Candy,
\newblock {\em Bayesian signal processing: classical, modern, and particle filtering methods}, vol.~54,
\newblock John Wiley \& Sons, 2016.

\bibitem{gilks1995}
W.~R. Gilks, S.~Richardson, and D.~Spiegelhalter,
\newblock {\em Markov Chain Monte Carlo in Practice},
\newblock Taylor \& Francis, 1995.

\bibitem{neal2011}
R.~M Neal,
\newblock ``{MCMC} using {H}amiltonian dynamics,''
\newblock in {\em Handbook of {M}arkov Chain {M}onte {C}arlo}, S.~Brooks, A.~Gelman, G.~L. Jones, and X.~Meng, Eds., chapter~5, pp. 113--162. Chapman and Hall/CRC, Boca Raton, USA, 2011.

\bibitem{girolami2011}
M.~Girolami and B.~Calderhead,
\newblock ``Riemann manifold {L}angevin and {H}amiltonian {M}onte {C}arlo methods,''
\newblock {\em J. Royal Statist. Society}, vol. 73, pp. 123 -- 214, Mar. 2011.

\bibitem{roberts2009}
G.~O. Roberts and J.~S. Rosenthal,
\newblock ``Examples of adaptive {MCMC},''
\newblock {\em J. Comput. and Graphical Statist.}, vol. 18, no. 2, pp. 349--367, 2009.

\bibitem{mahendran2012}
N.~Mahendran, Z.~Wang, F.~Hamze, and N.~D. Freitas,
\newblock ``Adaptive {MCMC} with {B}ayesian optimization,''
\newblock in {\em Proc. Int. Conf. Artificial Intell. and Statist.}, La Palma, Canary Islands, Apr. 2012, vol.~22, pp. 751--760.

\bibitem{elvira2015b}
V.~Elvira, L.~Martino, D.~Luengo, and M.~F. Bugallo,
\newblock ``Generalized multiple importance sampling,''
\newblock {\em Statist. Science}, vol. 34, Nov. 2015.

\bibitem{bugallo2017}
M.~F. {Bugallo}, V.~{Elvira}, L.~{Martino}, D.~{Luengo}, J.~{Miguez}, and P.~M. {Djuric},
\newblock ``Adaptive importance sampling: the past, the present, and the future,''
\newblock {\em IEEE Signal Process. Magazine}, vol. 34, no. 4, pp. 60--79, Jul. 2017.

\bibitem{cappe2004}
O.~Cappé, A.~Guillin, J.~M. Marin, and C.~P Robert,
\newblock ``Population {M}onte {C}arlo,''
\newblock {\em J. Computat. and Graphical Statist.}, vol. 13, no. 4, pp. 907--929, 2004.

\bibitem{elvira2017}
V.~{Elvira}, L.~{Martino}, D.~{Luengo}, and M.~F. {Bugallo},
\newblock ``Population {M}onte {C}arlo schemes with reduced path degeneracy,''
\newblock in {\em IEEE Int. Workshop on Computat. Adv. in Multi-Sensor Adaptive Process.}, Curacao, Dutch Antilles, Dec. 2017, pp. 1--5.

\bibitem{elvira2019}
V.~{Elvira} and {\'E}.~{Chouzenoux},
\newblock ``Langevin-based strategy for efficient proposal adaptation in population {M}onte {C}arlo,''
\newblock in {\em Proc. IEEE Int. Conf. Acoust., Speech and Signal Process.}, Brighton, UK, May 2019, pp. 5077--5081.

\bibitem{elvira2015}
V.~{Elvira}, L.~{Martino}, D.~{Luengo}, and J.~{Corander},
\newblock ``A gradient adaptive population importance sampler,''
\newblock in {\em Proc. IEEE Int. Conf. Acoust., Speech and Signal Process.}, Brisbane, Australia, Apr. 2015, pp. 4075--4079.

\bibitem{laham2019}
Y.~{El-Laham}, P.~M. {Djurić}, and M.~F. {Bugallo},
\newblock ``A variational adaptive population importance sampler,''
\newblock in {\em IEEE Int. Conf. on Acoust., Speech and Signal Process.}, Brighton, UK, May 2019, pp. 5052--5056.

\bibitem{elvira2017b}
V.~Elvira, L.~Martino, D.~Luengo, and M.~F. Bugallo,
\newblock ``Improving population monte carlo: alternative weighting and resampling schemes,''
\newblock {\em Signal Process.}, vol. 131, pp. 77--91, 2017.

\bibitem{elvira2022}
V.~Elvira and É. Chouzenoux,
\newblock ``Optimized population {M}onte {C}arlo,''
\newblock {\em IEEE Trans. Signal Process.}, vol. 70, pp. 2489--2501, May 2022.

\bibitem{elvira2023}
V.~Elvira, É. Chouzenoux, Ö.~D. Akyildiz, and L.~Martino,
\newblock ``Gradient-based adaptive importance samplers,''
\newblock {\em J. Franklin Inst.}, vol. 360, no. 13, pp. 9490--9514, 2023.

\bibitem{cappe2008}
O.~Cappé, R.~Douc, A.~Guillin, J.~M. Marin, and C.~P. Robert,
\newblock ``Adaptive importance sampling in general mixture classes,''
\newblock {\em Statist. Comput.}, vol. 18, pp. 447–459, 2008.

\bibitem{martino2017}
L.~Martino, V.~Elvira, D.~Luengo, and J.~Corander,
\newblock ``Layered adaptive importance sampling,''
\newblock {\em Statist. Comput.}, vol. 27, no. 3, pp. 599–623, 2017.

\bibitem{mousavi2021}
A.~Mousavi, R.~Monsefi, and V.~Elvira,
\newblock ``Hamiltonian adaptive importance sampling,''
\newblock {\em IEEE Signal Process. Letters}, vol. 28, pp. 713--717, Mar. 2021.

\bibitem{llorente2022}
F.~Llorente, E.~Curbelo, L.~Martino, V.~Elvira, and D.~Delgado,
\newblock ``{MCMC}‐driven importance samplers,''
\newblock {\em Appl. Math. Model.}, vol. 111, pp. 310--331, 2022.

\bibitem{cornuet2012}
J.~M. Cornuet, J.~M. Marin, A.~Mira, and C.~P. Robert,
\newblock ``Adaptive multiple importance sampling,''
\newblock {\em Scandinavian J. Statist.}, vol. 39, no. 4, pp. 798–812, Dec. 2012.

\bibitem{rezende2015}
D.~J. Rezende and S.~Mohamed,
\newblock ``Variational inference with normalizing flows,''
\newblock in {\em Pro. Int. Conf. Machine Learning}, Jul. 2015, p. 1530–1538.

\bibitem{dinh2017}
L.~Dinh, J.~Sohl-Dickstein, and S.~Bengio,
\newblock ``Density estimation using real {NVP},''
\newblock in {\em Proc. Int. Conf. Learning Representations}, Toulon, France, Apr. 2017.

\bibitem{papamakarios2017}
G.~{Papamakarios}, T.~{Pavlakou}, and I.~{Murray},
\newblock ``Masked autoregressive flow for density estimation,''
\newblock in {\em Proc. Adv. Neural Info. Process. Syst.}, Long Beach, CA, {USA}, Dec. 2017, pp. 2338--2347.

\bibitem{kingma2016}
D.~P. Kingma, T.~Salimans, R.~Jozefowicz, X.~Chen, I.~Sutskever, and M.~Welling,
\newblock ``Improved variational inference with inverse autoregressive flow,''
\newblock in {\em Proc. Adv. Neural Info. Process. Syst.}, Barcelona, Spain, Dec. 2016, pp. 4736--4744.

\bibitem{germain2015}
M.~Germain, K.l Gregor, I.~Murray, and H.~Larochelle,
\newblock ``{MADE}: {M}asked autoencoder for distribution estimation,''
\newblock in {\em Proc. Int. Conf. Machine Learning}, Lille, France, Jul. 2015, pp. 881--889.

\bibitem{trippe2018}
B.~L. Trippe and R.~E. {Turner},
\newblock ``Conditional density estimation with {B}ayesian normalising flows,''
\newblock in {\em Proc. Workshop on Bayesian Deep Learning, Adv. Neural Info. Process. Syst.}, Long Beach, CA, USA, Dec. 2017.

\bibitem{winkler2019}
C.~{Winkler}, D.~{Worrall}, E.~{Hoogeboom}, and M.~{Welling},
\newblock ``Learning likelihoods with conditional normalizing flows,''
\newblock {\em ArXiv e-prints: arXiv 1912.00042}, Nov. 2019.

\bibitem{lu2020}
Y.~Lu and B.~Huang,
\newblock ``Structured output learning with conditional generative flows,''
\newblock in {\em Proc. {AAAI} Conf. Artificial Intell.}, New York, NY, USA, Feb. 2020, pp. 5005--5012.

\bibitem{midgley2023}
L.~I. Midgley, V.~Stimper, G.~N.~C. Simm, B.~Sch{\"o}lkopf, and J.~M. Hern{\'a}ndez-Lobato,
\newblock ``Flow annealed importance sampling bootstrap,''
\newblock in {\em Proc. Int. Conf. Learning Representations}, Kigali, Rwanda, May 2023.

\bibitem{parno2018}
M.~D. Parno and Y.~M. Marzouk,
\newblock ``Transport map accelerated {M}arkov {C}hain {M}onte {C}arlo,''
\newblock {\em SIAM/ASA J. Uncertainty Quantification}, vol. 6, no. 2, pp. 645--682, 2018.

\bibitem{arbel2021}
M.~Arbel, A.~G. D.~G. Matthews, and A.~Doucet,
\newblock ``Annealed flow transport {M}onte {C}arlo,''
\newblock in {\em Proc. Int. Conf. Machine Learning}, Virtual, Jul. 2021.

\bibitem{gabrie2022}
M.~Gabrié, G.~M. Rotskoff, and E.~Vanden-Eijnden,
\newblock ``Adaptive {M}onte {C}arlo augmented with normalizing flows,''
\newblock {\em Proc. National Academy of Sciences}, vol. 119, no. 10, pp. e2109420119, 2022.

\bibitem{grumitt2022}
R.~D.~P. Grumitt, B.~Dai, and U.~Seljak,
\newblock ``Deterministic {L}angevin {M}onte {C}arlo with normalizing flows for {B}ayesian inference,''
\newblock in {\em Proc. Adv. Neural Info. Process. Syst.}, New Orleans, LA, USA, Dec. 2022.

\bibitem{roberts2002}
G.~O. Roberts and O.~Stramer,
\newblock ``Langevin diffusions and {M}etropolis-{H}astings algorithms,''
\newblock {\em Methodology and Comput. Appl. Probab.}, vol. 4, no. 4, pp. 337--357, Dec 2002.

\bibitem{duane1987}
S.~Duane, A.~D. Kennedy, B.~J. Pendleton, and D.~Roweth,
\newblock ``Hybrid {M}onte {C}arlo,''
\newblock {\em Phys. Lett. B}, vol. 195, no. 2, pp. 216 -- 222, 1987.

\bibitem{hoffman2013}
M.~D. Hoffman, D.~M. Blei, C.~Wang, and J.~Paisley,
\newblock ``Stochastic variational inference,''
\newblock {\em J. Machine Learning Research}, vol. 14, pp. 1303--1347, 2013.

\bibitem{glorot2010}
X.~Glorot and Y.~Bengio,
\newblock ``Understanding the difficulty of training deep feedforward neural networks,''
\newblock in {\em Proc. Int. Conf. Artificial Intell. and Statist.}, Sardinia, Italy, May 2010, pp. 249--256.

\end{thebibliography}

\IEEEpeerreviewmaketitle

\end{document}